\documentclass[10pt]{emulateapj}
\usepackage{apjfonts}

\usepackage{hyperref}
\usepackage{amssymb,amsmath}
\usepackage{color}

\newcommand{\degree}{\ensuremath{^\circ}}

\newcommand{\msun}{\ensuremath{\rm{M}_\odot}}
\newcommand{\rsun}{\ensuremath{\rm{R}_\odot}}

\newcommand{\kms}{\ensuremath{\rm{km \: s}^{-1}}}

\newcommand{\swift}{{\it Swift}}

\newcommand{\sneia}{SNe~Ia}
\newcommand{\snia}{SN~Ia}

\newcommand{\powerfittoall}{\ensuremath{ 2456998.88^{+ 0.11 }_{- 0.10 } }} 
\newcommand{\powerfittrise}{\ensuremath{ 16.94^{+ 0.11 }_{- 0.10 } }} 
\newcommand{\discoveryafter}{\ensuremath{ 2.23^{+ 0.11 }_{- 0.10 } }}

\newcommand{\velbesterror}{\ensuremath{ 2456996.93^{+ 0.81 }_{- 0.84 } }} 
\newcommand{\veldiffbesterror}{\ensuremath{ 1.93^{+ 0.81 }_{- 0.85 } }} 

\newcommand{\radiusconstraintab}{$0.34$ \rsun}
\newcommand{\radiusconstraint}{$0.34-11$ \rsun}
\newcommand{\primaryradius}{$0.5-6$ \rsun}

\begin{document}

\title{The Young and Bright Type Ia Supernova ASASSN-14lp: Discovery, Early-Time Observations, First-Light Time, Distance to NGC 4666, and Progenitor Constraints}
\shorttitle{ASASSN-14lp}
\shortauthors{Shappee et al.}	

\author{
{B.~J.~Shappee}\altaffilmark{1,2}, 
{A.~L.~Piro}\altaffilmark{1},
{T. W.-S.~Holoien}\altaffilmark{3},
{J.~L.~Prieto}\altaffilmark{4,5},
{C.~Contreras}\altaffilmark{6,7},
{K.~Itagaki}\altaffilmark{9},
{C.~R.~Burns}\altaffilmark{1},
{C.~S.~Kochanek}\altaffilmark{3,10},
{K.~Z.~Stanek}\altaffilmark{3,10},
{E. Alper}\altaffilmark{11},
{U.~Basu}\altaffilmark{3,12},
{J.~F.~Beacom}\altaffilmark{3,10,13},
{D.~Bersier}\altaffilmark{14},
{J.~Brimacombe}\altaffilmark{15},
{E.~Conseil}\altaffilmark{16},
{A.~B.~Danilet}\altaffilmark{13},
{Subo~Dong}\altaffilmark{17}, 
{E.~Falco}\altaffilmark{18},
{D.~Grupe}\altaffilmark{19},
{E.~Y.~Hsiao}\altaffilmark{6,7,8},
{S.~Kiyota}\altaffilmark{20},
{N.~Morrell}\altaffilmark{6},
{J.~Nicolas}\altaffilmark{21},
{M.~M.~Phillips}\altaffilmark{6},
{G.~Pojmanski}\altaffilmark{22},
{G.~Simonian}\altaffilmark{3},
{M.~Stritzinger}\altaffilmark{6,7},
{D.~M.~Szczygie{\l}}\altaffilmark{22},
{F.~Taddia}\altaffilmark{23},
{T.~A.~Thompson}\altaffilmark{3,10},
{J.~Thorstensen}\altaffilmark{11},
{M.~R.~Wagner}\altaffilmark{3,24},
{P.~R.~Wo\'zniak}\altaffilmark{25}
}

\email{bshappee@obs.carnegiescience.edu}

\altaffiltext{1}{Carnegie Observatories, 813 Santa Barbara Street, Pasadena, CA 91101, USA}
\altaffiltext{2}{Hubble, Carnegie-Princeton Fellow}
\altaffiltext{3}{Department of Astronomy, The Ohio State University, 140 West 18th Avenue, Columbus, OH 43210, USA}
\altaffiltext{4}{Nucleo de Astronomia de la Facultad de Ingenieria, Universidad Diego Portales, Av. Ejercito 441, Santiago, Chile}
\altaffiltext{5}{Millennium Institute of Astrophysics, Santiago, Chile}
\altaffiltext{6}{Carnegie Observatories, Las Campanas Observatory, Colina El Pino, Casilla 601, La Serena, Chile}
\altaffiltext{7}{Department of Physics and Astronomy, Aarhus University, Ny Munkegade 120, DK-8000 Aarhus C, Denmark}
\altaffiltext{8}{Department of Physics, Florida State University, Tallahassee, FL 32306, USA}

\altaffiltext{9}{Itagaki Astronomical Observatory, Teppo-cho, Yamagata 990-2492, Japan}
\altaffiltext{10}{Center for Cosmology and AstroParticle Physics (CCAPP), The Ohio State University, 191 W.\ Woodruff Ave., Columbus, OH 43210, USA}
\altaffiltext{11}{Department of Physics and Astronomy 6127 Wilder Laboratory, Dartmouth College Hanover, NH 03755-3528, USA}
\altaffiltext{12}{Grove City High School, 4665 Hoover Road, Grove City, OH 43123, USA}
\altaffiltext{13}{Department of Physics, The Ohio State University, 191 W.\ Woodruff Ave., Columbus, OH 43210, USA}
\altaffiltext{14}{Astrophysics Research Institute, Liverpool John Moores University, 146 Brownlow Hill, Liverpool L3 5RF, UK}
\altaffiltext{15}{Coral Towers Observatory, Cairns, Queensland 4870, Australia}
\altaffiltext{16}{AFOEV (Association Fran\c caise des Observateurs d'Etoiles Variables), Observatoire de Strasbourg 11, rue de l'Universit\'e, 67000 Strasbourg, France}
\altaffiltext{17}{Kavli Institute for Astronomy and Astrophysics, Peking University, Yi He Yuan Road 5, Hai Dian District, Beijing 100871, China}
\altaffiltext{18}{Harvard-Smithsonian Center for Astrophysics, 60 Garden Street, Cambridge, MA 02138, USA}
\altaffiltext{19}{Department of Earth and Space Science, Morehead State University, 235 Martindale Dr., Morehead, KY 40351, USA}
\altaffiltext{20}{Variable Stars Observers League in Japan (VSOLJ), 7-1 Kitahatsutomi, Kamagaya 273-0126, Japan}
\altaffiltext{21}{Groupe SNAUDE, France}
\altaffiltext{22}{Warsaw University Astronomical Observatory, Al. Ujazdowskie 4, 00-478 Warsaw, Poland}
\altaffiltext{23}{Department of Astronomy, The Oskar Klein Centre, Stockholm University, AlbaNova, 10691 Stockholm, Sweden}
\altaffiltext{24}{Large Binocular Telescope Observatory, University of Arizona, 933 N Cherry Avenue, Tucson, AZ 85721, USA}
\altaffiltext{25}{Los Alamos National Laboratory, Los Alamos, New Mexico 87545, USA}

\date{\today}

\begin{abstract}

On 2014 Dec.~9.61, the All-Sky Automated Survey for SuperNovae (ASAS-SN or ``Assassin'') discovered ASASSN-14lp just $\sim2$ days after first light using a global array of 14-cm diameter telescopes. ASASSN-14lp went on to become a bright supernova ($V = 11.94$ mag), second only to SN 2014J for the year.  We present prediscovery photometry (with a detection less than a day after first light) and ultraviolet through near-infrared photometric and spectroscopic data covering the rise and fall of ASASSN-14lp for more than 100 days.  We find that ASASSN-14lp had a broad light curve ($\Delta m_{15}(B) = 0.80  \pm  0.05$), a $B$-band maximum at $2457015.82 \pm 0.03$, a rise time of $\powerfittrise$ days, and moderate host--galaxy extinction ($E(B-V)_{\textrm{host}} = 0.33  \pm  0.06$).  Using ASASSN-14lp we derive a distance modulus for NGC 4666 of $\mu = 30.8  \pm  0.2$ corresponding to a distance of  $14.7 \pm 1.5$ Mpc.  However, adding ASASSN-14lp to the calibrating sample of Type Ia supernovae still requires an independent distance to the host galaxy.  Finally, using our early-time photometric and spectroscopic observations, we rule out red giant secondaries and, assuming a favorable viewing angle and explosion time, any non-degenerate companion larger than \radiusconstraintab.

\end{abstract}

\section{Introduction}
\label{sec:introduc}

Even though Type Ia supernovae (\sneia{}) have been used to discover the accelerating expansion of the universe \citep{riess98, perlmutter99}, the nature of their progenitor systems remains unknown (for a review see \citealp{maoz14}). It is generally accepted that \sneia{} involve the thermonuclear explosion of a C/O white dwarf (WD), but basic details of this picture, such as the nature of the binary companion and the sequence of events leading to the SN explosion, are still hotly debated.  Recently, significant observational (e.g., \citealp{gilfanov10, li11, horesh12, schaefer12, shappee13}) and theoretical progress (e.g.,~\citealp{kasen10, seitenzahl13, shappee13b, piro14, scalzo14}) has been made to constrain \sneia{} progenitor systems, but a clear picture has yet to emerge. 

Detailed, early-time observations are one of the more promising avenues to further constrain the progenitors of \sneia.  Statistical studies of the rise times ($t_{\textrm{rise}}$) of \sneia{} have a long history \citep{riess99, conley06, strovink07, hayden10, hayden10b, ganeshalingam11, gonzalezgaitan12, firth15}, but it has been the in-depth, early-time observations of nearby \sneia{} that have improved our understanding of the progenitor systems in recent years \citep{foley12, nugent11b, bloom12, silverman12, zheng13, yamanaka14, zheng14, siverd15, goobar15, olling15}.  The most notable example is SN~2011fe, which was discovered only hours after first light ($t_{\textrm{first}}$; \citealp{nugent11}) by the Palomar Transient Factory \citep{law09} at 6.4 Mpc \citep{shappee11}. \citet{bloom12} used a non-detection of SN~2011fe just 4 hours after $t_{\textrm{first}}$ to strongly constrain the radius of the primary ($R_* \lesssim 0.02$\rsun) and the radius of a companion ($R_\textrm{c} \lesssim 0.1$\rsun).  These constraints rule out all commonly considered non-degenerate companions and show directly, for the first time, that the exploding star for SN~2011fe was a degenerate object.  However, there might be multiple channels for producing normal SNe Ia (e.g., \citealp{maguire13}), highlighting the need to build a larger sample of well-constrained \sneia.

With the addition of ASASSN-14lp there are now 4 \snia{} with a clear detection within 1 day of first light.  
ASASSN-14lp was discovered just \discoveryafter{} days after first light and we present multiple pre-discovery detections.  In Section~\ref{sec:Obs}, we describe our discovery and observations.  In Section~\ref{sec:Distance}, we fit the light curve of ASASSN-14lp and determine the distance to its host galaxy. In Section~\ref{sec:constraints}, we use the early-time observations to constrain $t_{\textrm{first}}$ and the explosion time ($t_{\textrm{exp}}$). While $t_{\textrm{exp}}$ and $t_{\textrm{first}}$ have sometimes been used interchangeably in the literature, differences arises because of a possible dark phase between the explosion and when the SN first starts to optically brighten \citep{piro14}. Finally, in Section~\ref{sec:progenitor}, we combine all these results to constrain the radii of the exploding star and any non-degenerate companion.  We summarize our results in Section~\ref{sec:conclusion}.

\section{Discovery and Observations}
\label{sec:Obs}

The All-Sky Automated Survey for SuperNovae (ASAS-SN\footnote{\url{http://www.astronomy.ohio-state.edu/~assassin/index.shtml}} or ``Assassin''; \citealp{shappee14}) scans the entire extragalactic sky in both the Northern and Southern hemispheres roughly once every 2--3 nights in the $V$ band to depths of 16.5-17.3 mag depending on lunation (Shappee et al. 2016 in prep). We discovered a new source ($V \sim 14.9$ mag) on 2014 Dec.~9.61 (JD $t_{\textrm{disc}}= 2457001.112$) at $\textrm{RA}=12^{\rm h}45^{\rm m}09.\!\!^{\rm{s}}10$  $\textrm{decl}=-00\degree27'32.\!\!''5$ (J2000).  This location is $7\farcs5$ East and $10\farcs5$ North (physical distance of $\sim 1$ kpc) of the center of NGC 4666 ($z=0.0051$; \citealp{meyer04}), which is a member of the Virgo Cluster Southern Extension \citep{karachentsev13}. There are no pre-explosion images of this site available in the Hubble Space Telescope (HST) Archive. 

We designated this new source ASASSN-14lp and released its coordinates less than 5 hours after the discovery images were taken \citep{holoien14ATELb}, which allowed rapid follow-up at multiple wavelenths (e.g., \citealp{thorstensenATEL14, foleyATel14}).  We will present the prediscovery and follow-up photometric observations in Section \ref{sec:phot} and the spectroscopic follow-up observations in Section \ref{sec:spec}.

\begin{deluxetable}{lrcc}
\tablewidth{210pt}
\tabletypesize{\footnotesize}
\tablecaption{Photometric Observations}
\tablehead{
\colhead{JD} &
\colhead{Band} &
\colhead{Magnitude} &
\colhead{Telescope} \\ 
\colhead{($-$2,450,000)} &
\colhead{} &
\colhead{}  &
\colhead{}  }
\startdata
7001.906 & $UVW2$ & 17.35(0.09) & Swift \\ 
7003.772 & $UVM2$ & 17.83(0.10) & Swift \\ 
7001.902 & $UVW1$ & 16.32(0.08) & Swift \\ 
7002.836 & $u$ & 15.20(0.02) & CSP \\ 
7001.239 & $g$ & 14.91(0.03) & LCOGT \\ 
6837.541 & $V$ & >17.32 & ASAS-SN/bf \\ 
7001.834 & $r$ & 14.31(0.01) & CSP \\ 
6998.279 & $clear$ & >17.91 & Itagaki \\ 
7001.837 & $i$ & 14.51(0.01) & CSP \\ 
\enddata \tablecomments{The \swift, and $V\--\textrm{band}$ photometry are calibrated in the Vega magnitude system. The unfiltered photometry in calibrated to $r$-band in the Vega magnitude system. The SDSS $u'$, $g'$, $r'$, and $i'\--\textrm{band}$ photometry are calibrated in the AB magnitude system. \textit{Only the first observation in each band is shown here to demonstrate its form and content. Table is included in its entirety as an ancillary file..}} 
\label{tab:phot} 
\end{deluxetable}

\begin{figure*}
	\centerline{
		\includegraphics[width=17cm]{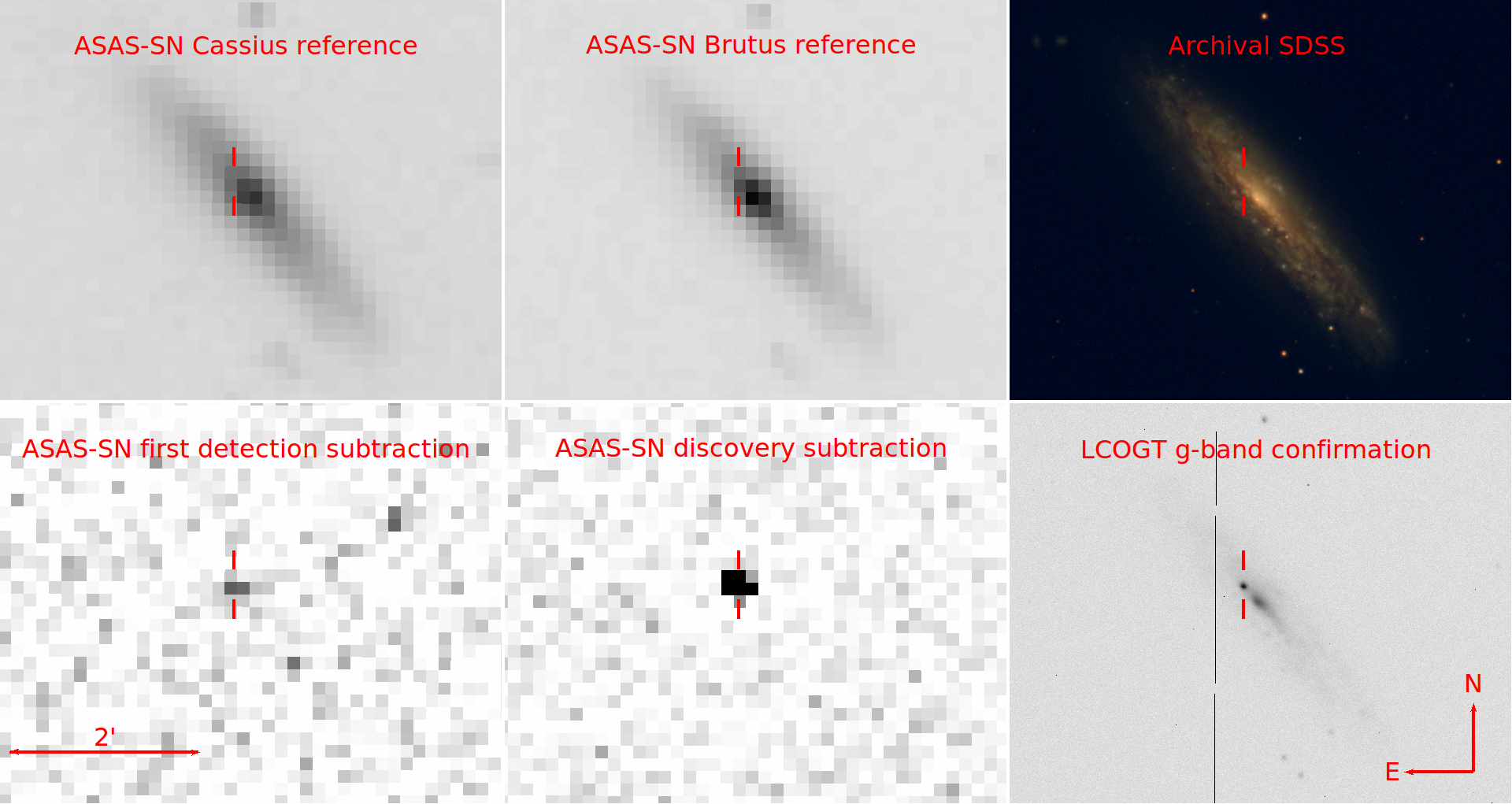}
	}
	\caption{Comparison between our ASAS-SN reference image from Cassius ({\bf top-left}), ASAS-SN first detection difference image from Cassius acquired 2014 Dec.~8.35 ({\bf bottom-left}), ASAS-SN reference image from Brutus ({\bf top-middle}), ASAS-SN discovery difference image from Brutus  acquired 2014 Dec.~9.61 ({\bf bottom-middle}), the archival Sloan Digital Sky Survey (SDSS) image ({\bf top-right}), our confirmation image from the Las Cumbres Observatory Global Telescope Network (LCOGT) 1-m at the MacDonald Observatory ({\bf bottom-right}).  All images are on the same angular scale and the position of ASASSN-14lp is marked.}
	\label{fig:ASASdisc}
\end{figure*}

\subsection{Photometric Observations}
\label{sec:phot}

\begin{figure*}
	\centerline{
		\includegraphics[height=17cm]{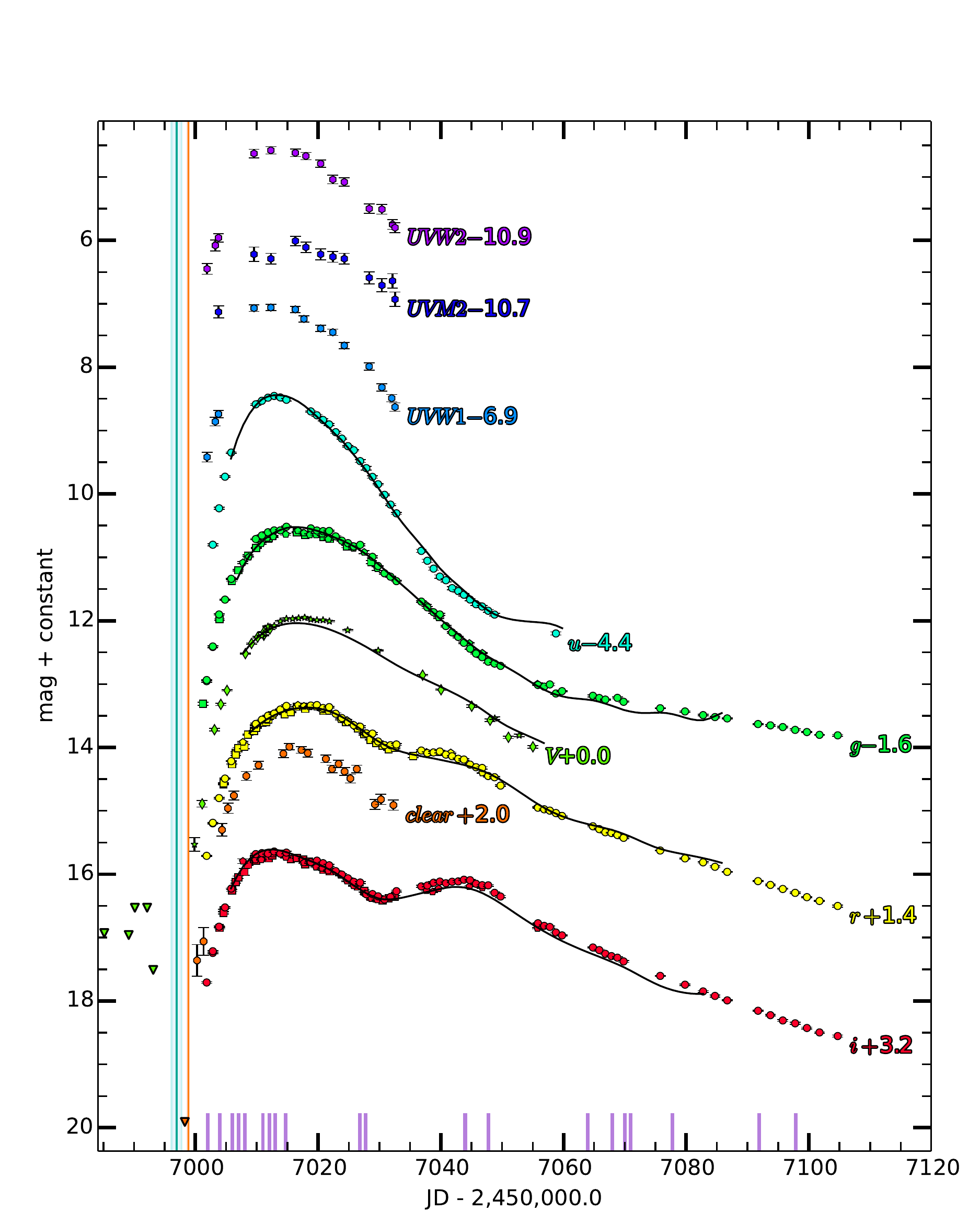}
	}
	\caption{Optical--UV photometric observations of ASASSN-14lp spanning $\sim105$ days from 2014 Dec.~8 through 2015 Mar.~23. Observations from ASAS-SN bd/bf cameras (diamonds/stars), CSPII (circles), K. Itagaki (octagons), LCOGT (squares), LT (pentagons), and \swift{} (hexagons) are shown with filled symbols. Downward-facing triangles represent upper limits.  Error bars are shown, but they are usually smaller than the points.  Black lines show the best-fit SNooPy light-curve models. The vertical orange line shows the best-fit $t_{\textrm{first}}$ (see Section~\ref{sec:lcfit}); its uncertainty is smaller than the width of the line.  The vertical turquoise line shows the best-fit $t_{\textrm{exp}}$ and its uncertainty (see Section~\ref{sec:velfit}).  Tick marks at the bottom of the plot denote the epochs of the spectra shown in Figure~\ref{fig:optical_spec}.}
	\label{fig:phot}
\end{figure*}

We have prediscovery images from two sources.  First, ASAS-SN has been observing NGC~4666 since January 2012 and had obtained more than 125 epochs at this location from both the northern unit (``Brutus'') on its fourth camera (bd) and from the southern unit (``Cassius'') on its second camera (bf) before the discovery of ASASSN-14lp.  Brutus obtained the discovery image on 2014 Dec.~9.61, with previous observations obtained 1.25, 7.98, 10.98, 11.97, 15.96, and 18.96 days before discovery (see Table \ref{tab:phot} and Figure~\ref{fig:phot} for details).  Images are processed by the fully automatic ASAS-SN pipeline (Shappee et al. 2016 in prep.) using the ISIS image subtraction package \citep{alard98, alard00}.  The reference image and first detection difference image for both bd and bf are shown in Figure~\ref{fig:ASASdisc}.  ASASSN-14lp is detectable in the bf image taken 1.25 days prior to its discovery, but this image was taken under poor conditions which prevented ASASSN-14lp from being flagged by our automated detection pipeline.  We performed aperture photometry on the subtracted images using the IRAF {\tt apphot} package and calibrated the results using the AAVSO Photometric All-Sky Survey (APASS; \citealp{henden15}).  The ASAS-SN detections and 3-sigma limits are presented in Table \ref{tab:phot}.

The second set of prediscovery images were obtained by K. Itagaki from 2014 Nov. 13 to 2015 Jan. 09 with a 0.30m f/8 reflector and an unfiltered CCD (KAF-1001E) at the Itagaki Astronomical Observatory, Japan.  The KAF-1001E CCD has a quantum efficiency curve\footnote{See Figure~2 in \url{http://www.kodak.com/ek/uploadedFiles/Content/Small_Business/Images_Sensor_Solutions/Datasheets(pdfs)/KAF-1001ELongSpecf}.} that reaches half its peak efficiency at $\sim 4100$ and $\sim 8900$ \AA.  We reduced the data using super-sky flats instead of twilight flats (which are unavailable).  Unfortunately, only 5 images were available from the period prior to our discovery of ASASSN-14lp.  For the image subtraction procedure we could use only one of the images for the reference image.  The other 4 either contain the SN, are needed to constrain the progenitor (see Section~\ref{sec:constraints}), or were taken in poor conditions. We calibrated the unfiltered CCD photometry using the APASS $r$-band magnitudes because $r$-band most closely matches the peak of the KAF-1001E detector quantum efficiency curve.  We matched 36 stars between the unfiltered images and APASS $r$-band catalog with magnitudes ranging from 10th to 13th mag.  The difference between the APASS magnitudes and instrumental magnitudes had a standard deviation of 0.094 mag (without any outlier rejection).  The magnitudes and the 3-sigma limits are presented in Table \ref{tab:phot}.

After the detection of ASASSN-14lp we began a multiwavelength, ground- and space-based follow-up campaign from the $X$-rays to the near-infrared (NIR) wavelengths.  \swift\ \citep{gehrels04} target-of-opportunity observations on both the UltraViolet/Optical Telescope (UVOT; \citealp{roming05}) and the $X$-ray Telescope (XRT; \citealp{burrows05}).  UVOT observations were obtained in $V$ (5468 \AA), $B$ (4392 \AA), $U$ (3465 \AA), $UVW1$ (2600 \AA), $UVM2$ (2246 \AA), and $UVW2$ (1928 \AA; \citealp{poole08}), spanning more than 30 days and covering the rise and fall of ASASSN-14lp. We used UVOTSOURCE to extract the source counts inside a $5\farcs0$ radius and from a sky region with a radius of $\sim40\farcs0$. However, since there is significant host galaxy contamination in the optical wavelengths, we present and analyze only the ultraviolet (UV) bands in this paper.  The UVOT Vega magnitudes presented in Table \ref{tab:phot} used the most recent UVOT calibrations \citep{poole08, breeveld10}. 

The XRT was operating in photon-counting mode \citep{hill04} during our observations. The data from all epochs were reduced and combined with the software tasks XRTPIPELINE and XSELECT to create a $0.3-10$ keV image at the location of ASASSN-14lp.  We found no point source at the location of ASASSN-14lp and the extended emission from NGC 4666 (XMM-08 in \citealp{dietrich06}) prevents us from placing any useful limits on the $X$-ray flux of the SN.

Optical ground-based images were obtained with the LCOGT 1-m network of telescopes ($gri$), the 2-m Liverpool telescope (LT) ($gri$), and the 40-inch Swope telescope ($ugri$) as part of the \textit{Carnegie Supernova Project II} (CSPII; \citealp{hsiao13}).  To remove the host-galaxy flux from the LCOGT and LT data, we used the HOTPANTS \footnote{\url{http://www.astro.washington.edu/users/becker/v2.0/hotpants.html}} package to subtract the SDSS archival $gri$ images of NGC~4666 from our new images.  The CSPII data was reduced in a similar manner to \citet{contreras10}, and we obtained satisfactory subtractions using the SDSS archival $ugri$ images of NGC~4666 as templates.

\subsection{Spectroscopic Observations}
\label{sec:spec}

We obtained an extensive sample of low- and medium-resolution optical spectra of ASASSN-14lp spanning more than 3 months between 2014 Dec.~10 and 2015 Mar.~16. Table~\ref{tab:spec} shows a summary of all the spectra, including date (UT+JD), telescope/instrument, wavelength range, spectral resolution, and exposure time. 

\begin{deluxetable*}{lcrrlcr}
\tabletypesize{\tiny}
\tablewidth{290pt}
\tablecaption{Spectroscopic Observations}
\tablehead{\colhead{} & 
\colhead{JD} &
\colhead{$t-t_{\textrm{first}}$} &
\colhead{$t-t_{Bmax}$} &
\colhead{} &
\colhead{Wavelength range} &
\colhead{Exposure} \\
\colhead{UT Date} &
\colhead{$-2$,400,000} &
\multicolumn{2}{c}{days} &
\colhead{Telescope/Instrument} &
\colhead{(\AA)} &
\colhead{(s)}   }
\startdata

2014 Dec 10.55 & 57002.048 &  +3.2  & -13.8  &  MDM2.4m/ModSpec        &  $4200\--7500$&       $240.0$ \\
2014 Dec 12.48 & 57003.976 &  +5.1  & -11.8  &  MDM2.4m/ModSpec        &  $4200\--7500$&       $600.0$ \\
2014 Dec 12.52 & 57004.021 &  +5.1  & -11.8  &  APO3.5m/DIS            &  $3500\--9600$&       $300.0$ \\
2014 Dec 14.52 & 57006.023 &  +7.1  & -9.8   &  FLWO1.5m/FAST          &  $3500\--7400$&       $180.0$ \\
2014 Dec 15.53 & 57007.025 &  +8.1  & -8.8   &  FLWO1.5m/FAST          &  $3500\--7400$&       $300.0$ \\
2014 Dec 15.56 & 57007.057 &  +8.2  & -8.8   &  MDM2.4m/ModSpec        &  $4200\--7500$&       $480.0$ \\
2014 Dec 16.54 & 57008.042 &  +9.2  & -7.8   &  MDM2.4m/ModSpec        &  $4200\--7500$&       $600.0$ \\
2014 Dec 19.51 & 57011.006 &  +12.1 &  -4.8  &  FLWO1.5m/FAST          &  $3500\--7400$&       $240.0$ \\
2014 Dec 20.54 & 57012.041 &  +13.2 &  -3.8  &  FLWO1.5m/FAST          &  $3500\--7400$&       $180.0$ \\
2014 Dec 21.54 & 57013.043 &  +14.2 &  -2.8  &  MDM2.4m/OSMOS          &  $4000\--6850$&       $300.0$ \\
2014 Dec 22.23 & 57014.730 &  +15.8 &  -1.1  &  NOT2.5m/ALFOSC         &  $3200\--9100$&       $900.0$ \\
2015 Jan  4.31 & 57026.814 &  +27.9 &  +11.0 &  Baade/IMACS            &  $3600\--9800$&       $300.0$ \\
2015 Jan  5.22 & 57027.724 &  +28.8 &  +11.9 &  NOT2.5m/ALFOSC         &  $3200\--9100$&       $720.0$ \\
2015 Jan 21.46 & 57043.963 &  +45.1 &  +28.1 &  LBT/MODS               & $3400\--10000$&       $1800.0$ \\
2015 Jan 21.49 & 57043.993 &  +45.1 &  +28.2 &  FLWO1.5m/FAST          &  $3500\--7400$&       $480.0$ \\
2015 Jan 25.29 & 57047.793 &  +48.9 &  +32.0 &  duPont/WFCCD           &  $3700\--9200$&       $400.0$ \\
2015 Feb 10.50 & 57063.997 &  +65.1 &  +48.2 &  MDM2.4m/OSMOS          &  $4000\--6850$&       $1200.0$ \\
2015 Feb 14.44 & 57067.941 &  +69.1 &  +52.1 &  FLWO1.5m/FAST          &  $3500\--7400$&       $600.0$ \\
2015 Feb 16.50 & 57070.002 &  +71.1 &  +54.2 &  LBT/MODS               & $3400\--10000$&       $900.0$ \\
2015 Feb 17.43 & 57070.933 &  +72.0 &  +55.1 &  FLWO1.5m/FAST          &  $3500\--7400$&       $600.0$ \\  
2015 Feb 24.25 & 57077.752 &  +78.9 &  +61.9 &  duPont/WFCCD           &  $3700\--9200$&       $300.0$ \\
2015 Mar 10.38 & 57091.879 &  +93.0 &  +76.1 &  MDM2.4m/OSMOS          &  $4000\--6850$&       $600.0$ \\
2015 Mar 16.34 & 57097.837 &  +98.9 &  +82.0 &  FLWO1.5m/FAST          &  $3500\--7400$&       $900.0$ 

\enddata 

\label{tab:spec}

\end{deluxetable*}

The single-slit spectra from the Modular Spectrograph (Modspec) mounted on the MDM Observatory Hiltner 2.4-m telescope, DIS on the APO 3.5-m telescope, the Ohio State Multi-Object Spectrograph (OSMOS; \citealp{martini11}) on the MDM Observatory Hiltner 2.4-m telescope, the Inamori-Magellan Areal Camera \& Spectrograph (IMACS; \citealp{dressler11}) on the Baade-Magellan~6.5-m telescope, the Andalucia Faint Object Spectrograph and Camera (ALFOSC) on the Nordic Optical 2.5-m Telescope (NOT), and the Wide Field Reimaging CCD Camera (WFCCD) on the du Pont 100-inch Telescope were all reduced with standard routines in the IRAF {\tt twodspec} and {\tt onedspec} packages.  The spectra obtained at the F. L. Whipple Observatory (FLWO) 1.5 m Tillinghast telescope using the FAST spectrograph \citep{fabricant98} were reduced using a combination of standard IRAF and custom IDL procedures \citep{matheson05}.  The spectra from the Multi-Object Double Spectrographs (MODS; \citealp{pogge10}) on the dual 8.4-m Large Binocular Telescope (LBT) on Mount Graham were reduced using a custom pipeline written in IDL\footnote{\url{http://www.astronomy.ohio-state.edu/MODS/Software/modsIDL/}}.  The reductions included bias subtraction, flat-fielding, 1-D spectral extraction, wavelength calibration using an arc lamp, and flux calibration using a spectroscopic standard usually taken the same night. We calibrate the spectra by extracting synthetic photometric magnitudes for each filter in the spectral range and then finding the best linear fit to the observed magnitudes interpolated to the spectral epoch. For spectra where only a single filter fit within the spectral range, the spectrum was scaled by a constant.  Finally, while no filter is completely inside the OSMOS spectral range, virtually all ($>98\%$) of the $g$-band transmission function lies in the observed spectral range, so we calibrate to this filter.  

\begin{figure*}
	\centerline{
		\includegraphics[width=15cm]{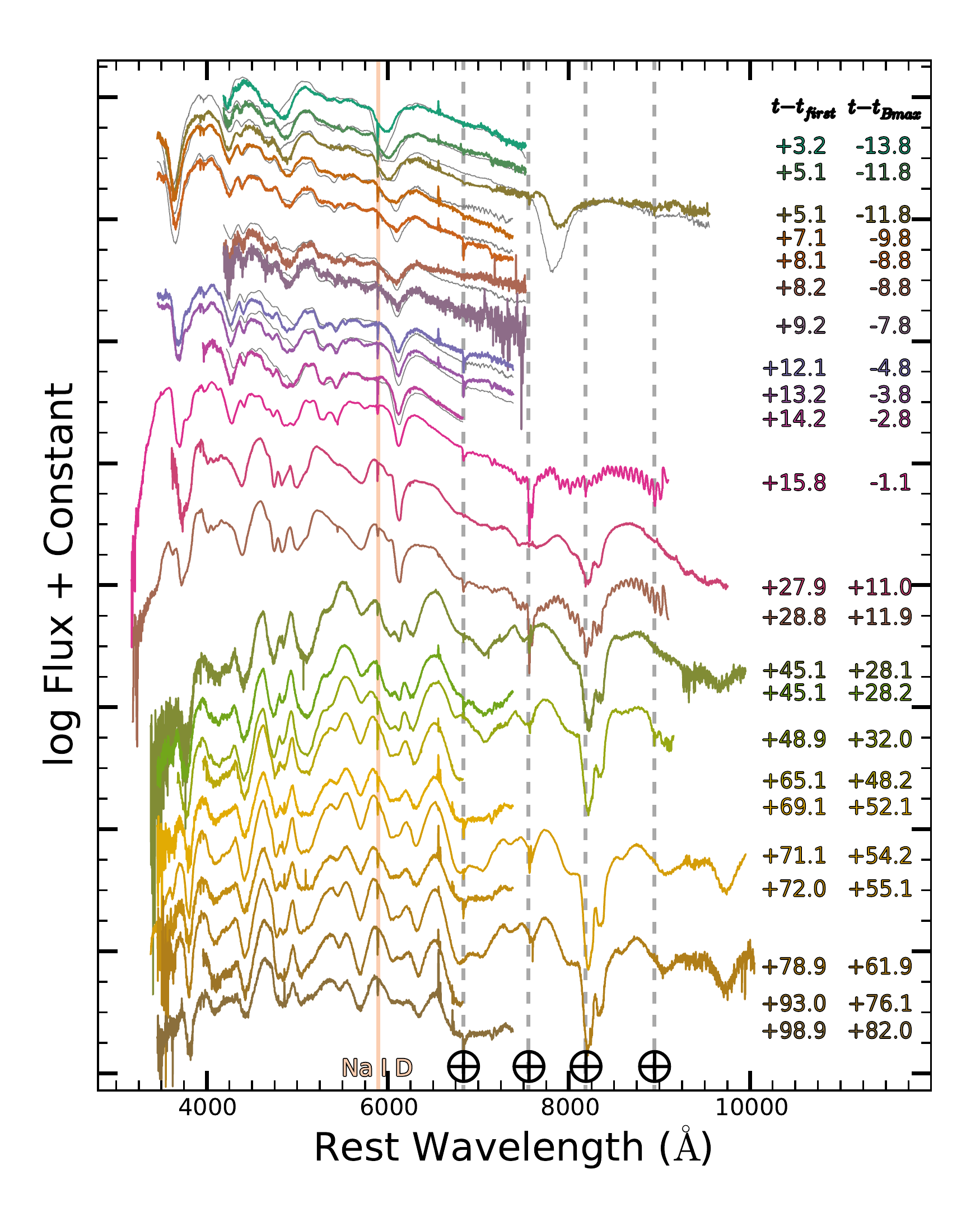}
	}
	\caption{Spectroscopic observations of ASASSN-14lp spanning from 2014 Dec.~10 -- 2015 Mar.~16. Phases relative to our estimate of $t_{\textrm{first}}$ and $t_{Bmax}$ are indicated. Details of the observations are given in Table~\ref{tab:spec}. Individual spectra are calibrated using broad-band photometry interpolated to the spectral epoch.  For the first 15 days after first light, spectra of SN~2009ig interpolated to the same epoch are shown in gray \citep{foley12}. Vertical dashed gray lines mark strong telluric features.  Vertical apricot line marks host-galaxy Na I D absorption. }
	\label{fig:optical_spec}
\end{figure*}

The calibrated spectra are shown in Figure~\ref{fig:optical_spec} with early-time spectra of SN~2009ig \citep{foley12} shown underneath for reference. Strong host-galaxy Na I D absorption is visible in all ASASSN-14lp spectra, which is promising for time-viability studies to constrain the presence of circumstellar material (e.g., \citealp{patat07, simon09}).

\begin{figure}
	\centerline{
		\includegraphics[height=13cm]{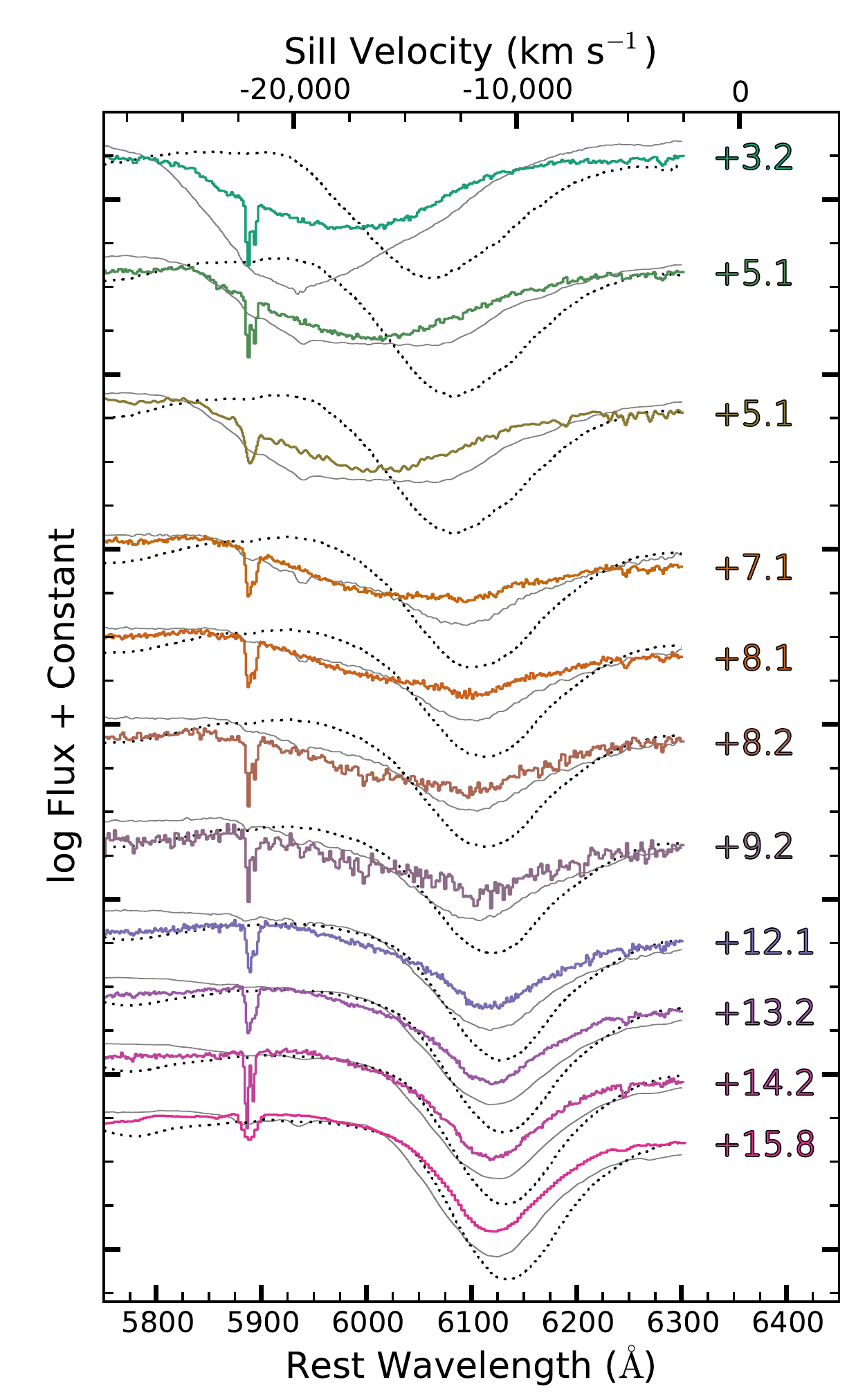}
	}
	\caption{The pre-maximum evolution of the region near the Si ${\textrm{II}} \lambda 6355$ feature.  Phases from our derived $t_{\textrm{first}}$ from Section~\ref{sec:lcfit} are indicated. The  \citet{foley12} spectra of SN~2009ig (solid gray) and the \citet{pereira13} spectra of SN~2011fe (dotted black), interpolated to the same epochs as ASASSN-14lp, are also shown. Like ASASSN-14lp, SN~2009ig also had a high-velocity Si II component that gives way to a lower-velocity component starting roughly 5 days after first light.}
	\label{fig:Si_spec}
\end{figure}

In Figure~\ref{fig:Si_spec}, we show the pre-max evolution of the spectroscopic region near the Si ${\textrm{II}} \lambda 6355$.  For comparison, we also show the spectra of the two other,  SN~2009ig \citet{foley12} and SN~2011fe \citep{pereira13}, that were both observed within 1 day of first light and had early-time spectroscopic observations.  These spectra were interpolated in time to the same epoch since $t_{\textrm{first}}$ as ASASSN-14lp.  While detailed modeling of these spectra is beyond the scope of this paper, it is interesting to note that a high velocity Si II component ($\sim 18,000$ \kms) is clearly visible in the ASASSN-14lp spectra at early times.  The Si II feature then becomes dominated by a lower-velocity component ($\sim 13,000$ \kms) starting 5 days after $t_{\textrm{first}}$.  This behavior is qualitatively similar to SN~2009ig but was not observed in SN~2011fe \citep{piro14}.

\section{Light-Curve Fit and Distance to NGC 4666}
\label{sec:Distance}

We used the CSPII light-curve fitting code SuperNovae in Object Oriented Python (SNooPy; \citealp{burns11}) to fit the light curve of ASASSN-14lp and to derive a distance to NGC 4666.  We employed SNooPy's {\it systematics} function, the spectral template from \citet{hsiao07}, and the ``Best-observed'' calibration (Fit~4 from Table~9 of \citealp{folatelli10}). We adopted a Galactic extinction of $E(B-V)_{\textrm{MW}} = 0.021$ mag \citep{schlegel98, schlafly11} and a R$_{V} = 3.1$ reddening law \citep{cardelli89}.  

The results of the SNooPy fit are shown in Table~\ref{tab:snoopy}.  The light curve of ASASSN-14lp is noteworthy because it is broad ($\Delta m_{15}(B) = 0.80  \pm  0.05$) but not outside the range for normal supernovae ($\Delta m_{15}(B) \sim 0.79-1.62$; see Figure 4 in \citealp{stritzinger11}).  Employing the relation in \citet{mazzali07}, the estimate of $\Delta m_{15}(B)$ yields a $^{56}$Ni mass of $0.806 \pm 0.013$ \msun{}, which is high but not unique among \sneia{} (e.g., Figure~1 in \citealp{piro14b}). The time of $B$-band maximum is JD $t_{Bmax} = 2457015.82 \pm 0.03$, $14.728$ days after discovery. We see that ASASSN-14lp suffered moderate host galaxy extinction, with $E(B-V)_{\textrm{host}} = 0.33  \pm  0.06$.   Finally, the derived distance modulus to NGC 4666 is $\mu = 30.8  \pm  0.2$.  This distance modulus implies a physical distance to NGC 4666 of $14.7 \pm 1.5$ Mpc.


\begin{deluxetable*}{lrccc}
\tablewidth{460pt}
\tabletypesize{\footnotesize}
\tablecaption{Derived ASASSN-14lp Properties}
\tablehead{
\colhead{$t_{Bmax}$} &
\colhead{$E(B-V)_{\textrm{host}}$} &
\colhead{$\Delta m_{15}$} &
\colhead{$\mu$} &
\colhead{$d$} \\ 
\colhead{($-$2,450,000)} &
\colhead{(mag)} &
\colhead{(mag)}  &
\colhead{(mag)}  &
\colhead{(Mpc)}  }
\startdata
$7015.82 \pm 0.03$ & $0.33  \pm  0.06$   &   $0.80  \pm  0.05$   &   $30.8  \pm 0.2$   &   $14.7 \pm 1.5$ \\


\enddata \tablecomments{The quoted errors include systematic errors as reported by SNooPy's {\it systematics} function \citep{burns11}.}
\label{tab:snoopy} 
\end{deluxetable*}

\section{First Light and Explosion Time Constraints}
\label{sec:constraints}

In this section we present constraints on $t_{\textrm{first}}$ for ASASSN-14lp from fits to the early-time light curve (Section \ref{sec:lcfit}), and from the early-time expansion velocity we constrain $t_{\textrm{exp}}$  (Section~\ref{sec:velfit}).

\subsection{Early-Time Light Curve Fit}
\label{sec:lcfit}

\begin{figure}
	\centerline{
		\includegraphics[height=13cm]{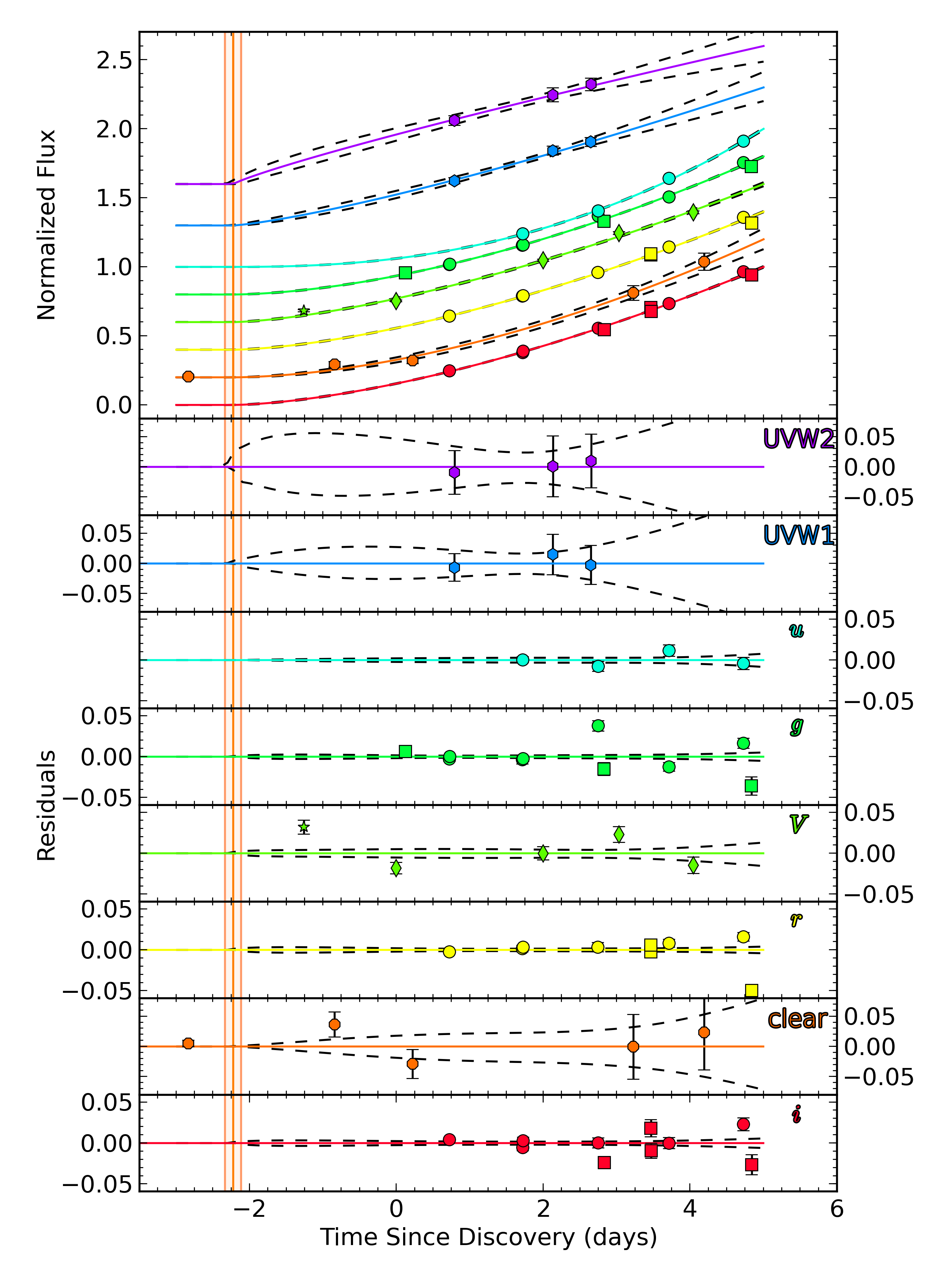}
	}
	\caption{Early-time light curves from 3 days before to 6 days after discovery with power-law fits. The colors and symbols are the same as in Figure~\ref{fig:phot}. Colored solid lines are the best-fit power laws while dashed black lines show each fit's 1-sigma uncertainties.  The vertical orange line shows the best-fit $t_{\textrm{first}}$ and its uncertainty (see Section~\ref{sec:lcfit}).  {\bf Top panel:}  Light curves normalized by their flux at 5 days after discovery with an added constant.  {\bf Bottom panels:} Residuals from the best-fit power laws for each individual filter.}
	\label{fig:earlyfit}
\end{figure}

To determine $t_{\textrm{first}}$, we first attempted to model the early-time ASASSN-14lp light curve with an expanding fireball model where flux is proportional to $(t-t_{\textrm{first}})^{2}$ .  However, this lead to unsatisfactory fits.  We then added more freedom to our model, fitting an independent power-law index ($\alpha_{i}$) for each band $\propto (t - t_{\textrm{first}})^{\alpha_i}$, but forcing them all to have the same $t_{\textrm{first}}$.  We used the {\tt emcee} package \citep{foreman13}, a Python-based implementation of the affine-invariant ensemble sampler for Markov chain Monte Carlo (MCMC), to perform the fit to each light curve. We excluded the \swift{} $UVM2$-band from this analysis because ASASSN-14lp was not clearly detected within  5 days of discovery. The best-fit power laws for each filter and their corresponding 1-sigma uncertainties are shown in Figure~\ref{fig:earlyfit}.   
We find that JD $t_{\textrm{first}} = \powerfittoall$, implying that we discovered ASASSN-14lp just \discoveryafter{} days after $t_{\textrm{first}}$ and that $t_{\textrm{rise}} = t_{Bmax} - t_{\textrm{first}} = \powerfittrise$ days.  The best-fit $\alpha$ for each band are presented in Table~\ref{tab:lcfit}.  The $g$, $V$, $r$, $clear$, and $i$ filters have best-fit $\alpha \sim 1.50-1.75$, but the $u$ filter requires a steeper rise.

\begin{deluxetable}{crrr}
\tablewidth{200pt}
\tabletypesize{\small}
\tablecaption{Fit Light Curve Parameters}
\tablehead{
\colhead{band} &
\colhead{$t_{\textrm{max}}$} &
\colhead{$m_{\textrm{max}}$} &
\colhead{$\alpha$} \\ 
\colhead{} &
\colhead{($-$2,450,000)} &
\colhead{(mag)}  &
\colhead{} } 
$UVW2$ & $7013.12\pm0.77$ & $15.61\pm0.03$ &  $ 1.45^{+ 0.46 }_{- 0.43 } $ \\ 
 $UVM2$ & $7016.52\pm1.03$ & $17.36\pm0.08$ & \nodata \\ 
$u$ & $7015.14\pm0.21$ & $12.84\pm0.03$ &  $ 2.38^{+ 0.05 }_{- 0.05 } $ \\ 
 $g$ & $7017.57\pm0.20$ & $12.15\pm0.02$ &  $ 1.70^{+ 0.04 }_{- 0.04 } $ \\ 
 $V$ & $7018.11\pm0.19$ & $11.94\pm0.01$ &  $ 1.50^{+ 0.05 }_{- 0.05 } $ \\ 
 $r$ & $7018.01\pm0.11$ & $11.94\pm0.01$ &  $ 1.57^{+ 0.04 }_{- 0.04 } $ \\ 
 $clear$ & $7016.67\pm0.21$ & $12.04\pm0.02$ &  $ 1.75^{+ 0.19 }_{- 0.16 } $ \\ 
 $i$ & $7015.02\pm0.22$ & $12.50\pm0.01$ &  $ 1.58^{+ 0.04 }_{- 0.04 } $ \\ 
 \enddata  
\label{tab:lcfit} 
\end{deluxetable}

The first photometric observations around 1 day after $t_{\textrm{first}}$ from both ASAS-SN and K. Itagaki are brighter than the simple power-law fits by about 3 and 1 sigma, respectively.  
These are weak detections, but they perhaps hint that a broken power-law fit would be a better description at these very early times, similar to SN 2013dy \citep{zheng13}. However, with only one photometric point in each band during the first two days after $t_{\textrm{first}}$, such a fit is not well-constrained.  Both of these points are fully accounted for when placing constraints on the progenitor system in Section~\ref{sec:progenitor}.

\begin{figure}
	\centerline{
		\includegraphics[width=8.5cm]{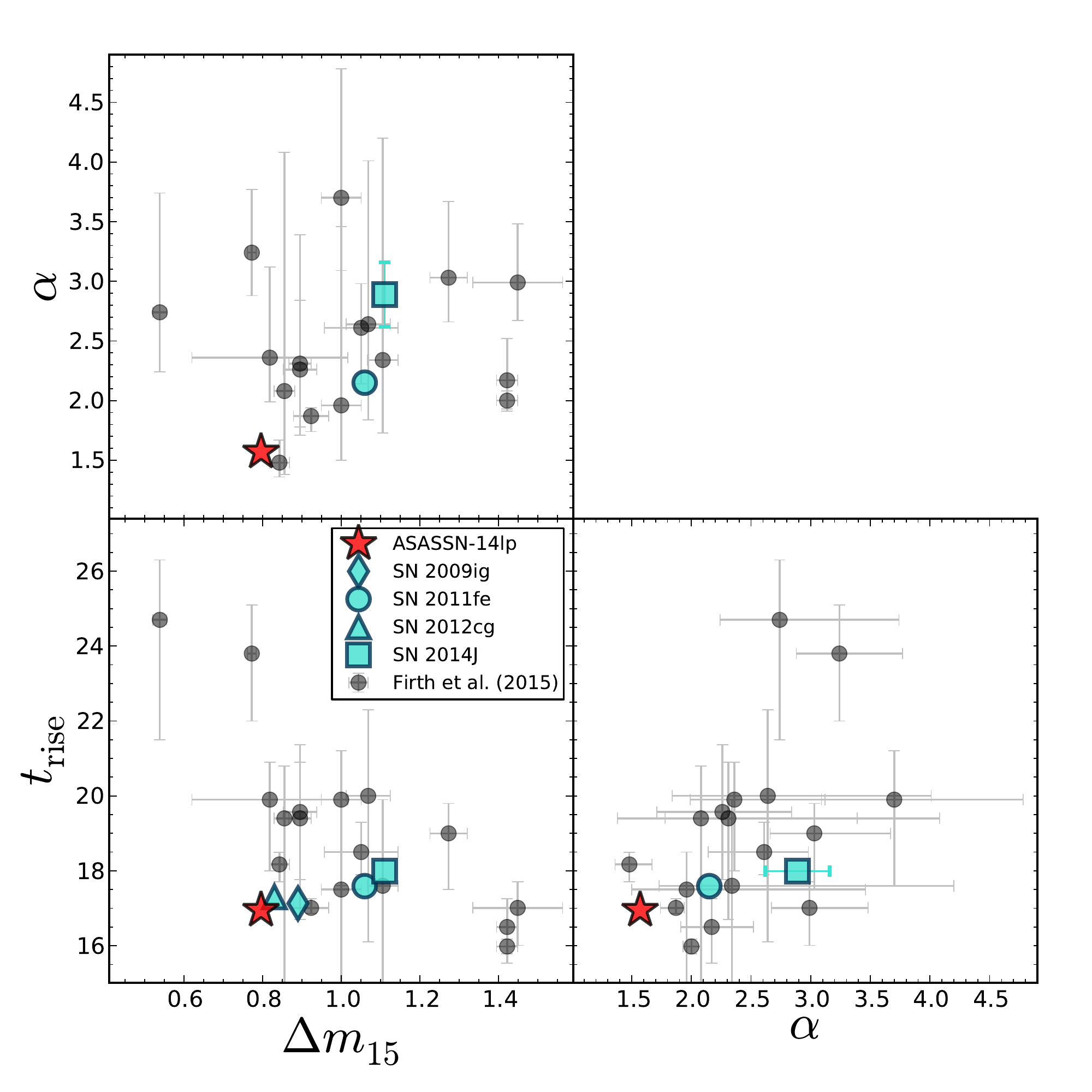}
	}
	\caption{Comparison between $t_{\textrm{rise}}$, $\Delta m_{15}(B)$, and $\alpha$ for ASASSN-14lp (red star), SN~2009ig (blue diamond; \citealp{foley12}), SN~2012cg (blue triangle; \citealp{silverman12}), and the \citet{firth15} collection of SNe (grey circles) which includes SN~2011fe (blue circle). Error bars for ASASSN-14lp, SN~2009ig, SN~2012cg, and SN~2011fe are smaller than the points. }
	\label{fig:compare}
\end{figure}

In Figure~\ref{fig:compare}, we compare $t_{\textrm{rise}}$, $\Delta m_{15}(B)$, and $\alpha$ for ASASSN-14lp against all available prior measurements: SN~2009ig \citep{foley12}, SN~2012cg \citep{silverman12}, SN~2014J \citep{zheng14, siverd15, marion15}, and the sample (including SN~2011fe) from \citet{firth15}.  We use Equation~5 of \citet{conley08} to translate the stretch reported in Table~4 of \citet{firth15} into $\Delta m_{15}(B)$.  We compare our $r$-band $\alpha$ to those reported by \citet{firth15}. ASASSN-14lp is on the edge of, but consistent with, these distributions.  In summary, ASASSN-14lp is normal, but it has a post-max light curve that is broad, an overall rise time that is short, and an early-time rise that is slow.  

\subsection{Early-time Expansion Velocity Fit}
\label{sec:velfit}

Following \citet{piro14}, we use the early-time spectroscopic observations to measure line velocities and then by assuming a theoretically motivated, time-dependent velocity evolution model we estimate $t_{\textrm{exp}}$.  We then compare $t_{\textrm{exp}}$, determined by fitting these early-time spectroscopic observations, and $t_{\textrm{first}}$, determined by fitting the early-time light curve, to constrain the existence of a possible dark phase.

First, we measured the velocities of the Ca ${\textrm{II}}$ H and K, and Si ${\textrm{II}} \lambda 6355$ absorption features for the spectra presented in Table~\ref{tab:spec} taken within 15 days of $t_{\textrm{first}}$.  We used the wavelength of the deepest absorption to determine the velocity.  Similar to \citet{piro14}, we assume an error of 500 km/s as a rough estimate of the typical measurement error \citep{parrent12}. As previously mentioned, the spectra in Figure~\ref{fig:Si_spec} show that there are two components to the Si ${\textrm{II}} \lambda 6355$ line: a high-velocity component which dominates the first $\sim 5$ days after $t_{\textrm{first}}$ and a low-velocity component which dominates afterwards.  Measuring the velocity of both Si ${\textrm{II}} \lambda 6355$ components at all epochs would require detailed modeling of the spectra, which is beyond the scope of this paper. Instead, we simply measured the deepest absorption of the dominant Si ${\textrm{II}} \lambda 6355$ component in each spectrum.

\begin{figure}
	\centerline{
		\includegraphics[width=8.5cm]{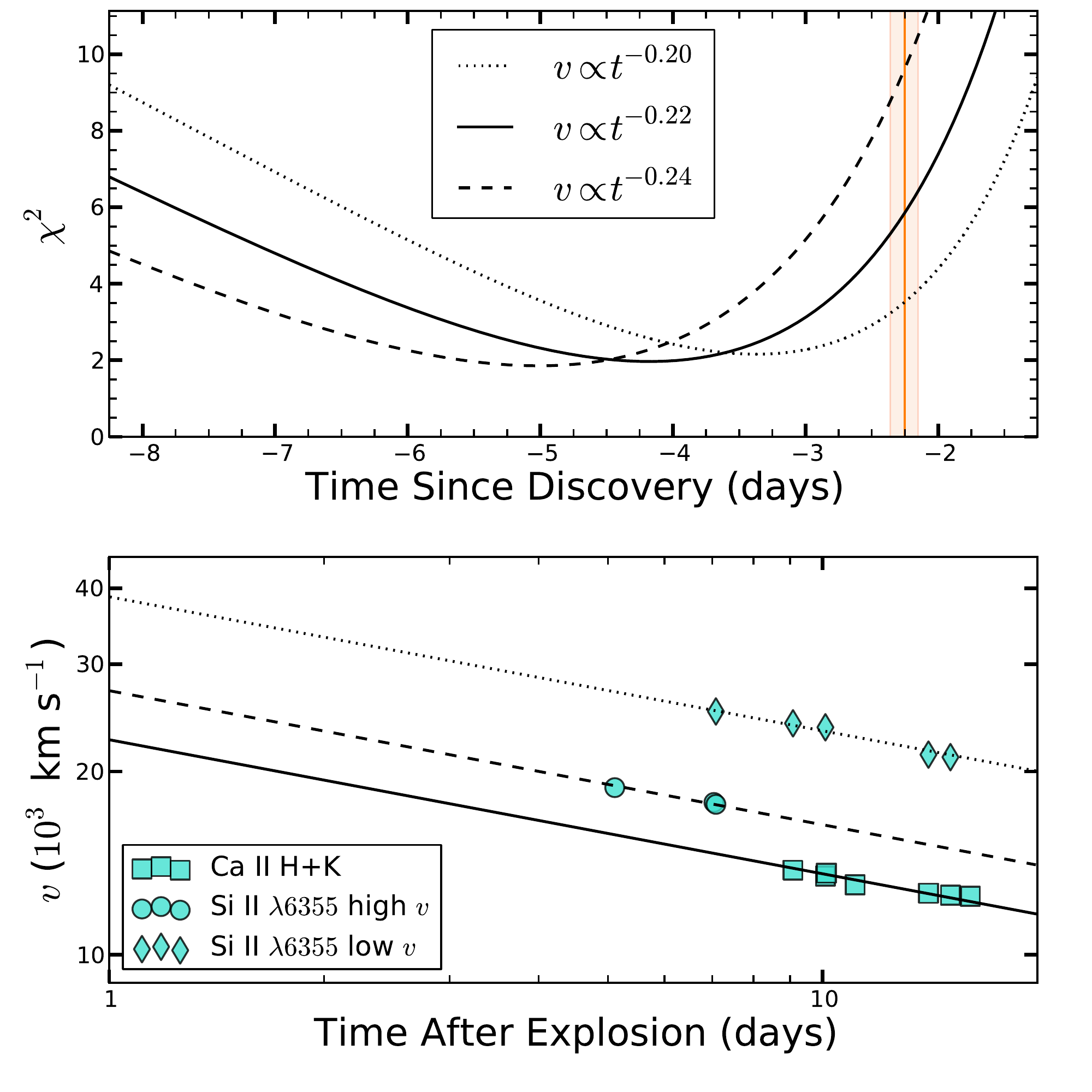}
	}
	\caption{Constraints on $t_{\textrm{exp}}$ from fitting the velocity evolution of the Ca ${\textrm{II}}$ H and K, and Si ${\textrm{II}} \lambda 6355$ following \citet{piro14}. {\bf Top panel:} The dotted, solid, and dashed lines show the $\chi^2$ for fit power laws with indices $-0.20$, $-0.22$, and $-0.24$, respectively, as a function of assumed explosion date.  The orange line and region show the best-fit $t_{\textrm{first}}$ and its 1-sigma error bars from Section~\ref{sec:lcfit}.  {\bf Bottom panel:}  Observed low-velocity Si ${\textrm{II}} \lambda 6355$ (squares), high-velocity Si ${\textrm{II}} \lambda 6355$ (circles), and Ca ${\textrm{II}}$ H and K (diamonds) with best-fit power-law velocity with $v \propto t^{-0.22}$.}
	\label{fig:velfit}
\end{figure}

We then fit these line velocities by assuming they have particular power-law evolution in time and varying the explosion time. The expected velocity evolution for the spectral lines of very young \sneia{} is a power law with time ($v \propto t^{-0.22}$) if the lines are generated at a constant opacity \citep{piro14}. The bottom panel of Figure~\ref{fig:velfit} shows the Ca ${\textrm{II}}$ H and K, and Si ${\textrm{II}} \lambda 6355$ velocities and their best fit assuming $v \propto t^{-0.22}$.  \citet{piro14} use the spread in the best fit explosion time between $v \propto t^{-0.20}$ and $v \propto t^{-0.24}$ as an estimate of the uncertainty. The top panel of Figure~\ref{fig:velfit} shows the goodness of fit, as a function of explosion time, for these three power laws: $v \propto t^{-0.20}$, $v \propto t^{-0.22}$, and $v \propto t^{-0.24}$.  This leads us to JD $t_{\textrm{exp}} = \velbesterror$ which is \veldiffbesterror{} days earlier than $t_{\textrm{first}}$.

The offset between $t_{\textrm{exp}}$ and $t_{\textrm{first}}$ is similar to that found by \citet{piro14} for SN~2009ig, which is spectroscopically (see Section~\ref{sec:spec}) and photometrically similar (SN~2009ig also had a broad light curve with $\Delta m_{15}(B) = 0.89$; \citealp{foley12}) to ASASSN-14lp.  There are three possible explanations for this difference:  1) The modeling method presented in \citet{piro14} does not accurately represent any \sneia, perhaps because of the assumption that the opacity near the photosphere is roughly constant or the velocity profile of the material is different.  Although possible, one would still need to explain why this method seemed to work well for SN~2011fe and SN~2012cg and not for SN~2009ig and ASASSN-14lp.  2) The modeling method presented in \citet{piro14} is fine in general, but something about the SN~2009ig and ASASSN-14lp explosions breaks its assumptions.  \citet{piro14} suggest that the discrepancy might be due to the explosions being asymmetric, with higher velocity material coming towards the observer \citep{maeda10}.  3) There is a dark time between when the explosion occurs and when optical brightening occurs for some SNe \citep{piro13}, and fitting a power law to the early-time light curve and extrapolating this to earlier times does not always accurately estimate explosion times.  Distinguishing between these three possibilities is beyond the scope of this paper, but it is an important question for early-time studies of \sneia because the presence of a dark time between explosion and first light can drastically alter inferences about the structure of the progenitor. To help investigate the underlying mechanism that causes these features, it will be useful to explore whether a larger sample of early-time light curves can be put into groups that share basic properties, as we have seen for SN~2009ig and ASASSN-14lp or SN~2011fe and SN~2012cg.

\section{Progenitor System Constraints}
\label{sec:progenitor}
To place constraints on the progenitor system of ASASSN-14lp, we first create absolute magnitude light curves from the two filters, ASAS-SN $V$ and K. Itagaki's clear, with very early (within 2 days of $t_{\textrm{first}}$) detections or upper limits. In this section we treat the clear filter as $r$, based on the calibrations in Section~\ref{sec:phot}.   One could use the distance modulus and reddening determined in Section~\ref{sec:Distance} to compute absolute magnitude light curves from our observed light curves, however, this would be circular because SNooPy assumed an absolute magnitude of ASASSN-14lp to derive its distance.  Instead, we scale our observed $V$- and $r$-band light curves to the expected absolute maximum given the observed value of $\Delta m_{15}(B)$ and the ``Best-observed'' calibration of \citet{folatelli10} for these filters (their Fits~7 and 10 from Table~9, respectively).  To do this we determine the peak magnitude ($m_\textrm{max}$) in each filter by fitting a quadratic to each filter's light curve over $\pm10$ days around $t_{Bmax}$.  The time of maximum ($t_{\textrm{max}}$) and $m_\textrm{max}$ for each band are presented in Table~\ref{tab:lcfit}.  We add the errors induced by the intrinsic scatter of \sneia{} ($\sigma_{\textrm{SN}}$), the intercept of the Phillips relation ($M_\textrm{X}$), the slope of the Phillips relation ($b_\textrm{X}$), and the estimate of $m_\textrm{max}$ in quadrature with the photometric errors.  This means that the photometric points in the right panel of Figure~\ref{fig:companion} have correlated errors.  For the upper limits, we take a conservative approach and add the quadrature of these errors to the upper limit.   We use the early-time absolute magnitude light curves of ASASSN-14lp to constrain its progenitor system in two ways. In Section~\ref{sec:companion}, we constrain the radius of a possible non-degenerate companion by placing limits on its interaction with the SN ejecta.  In Section~\ref{sec:primary}, we constrain the radius of the exploding star itself by placing limits on flux from its shock-heated envelope.

\subsection{Constraints on a Non-Degenerate Companion's Radius}
\label{sec:companion}

If the progenitor of an \snia{} is a WD accreting from a non-degenerate companion, then the SN ejecta will interact with the companion and potentially produce an observable signature in the rising light curve at early times. Such a signature is dependent on the viewing angle, with the strongest effect occurring when the companion lies along the line of sight between the observer and the SN. At a fixed viewing angle, this emission scales proportionally with $R_\textrm{c}$, which allows early time detections or upper limits to constrain the properties of the companion.

We apply the analytic models provided by \citet{kasen10} to place constraints on the radius of a potential companion to ASASSN-14lp. Since the explosion time is uncertain, we explore a variety of explosion times and their corresponding constraints on $R_\textrm{c}$. The general procedure is to choose an explosion time and then find the maximum radius companion that is consistent with the early-time data assuming the secondary filled its Roche-lobe. 

\begin{figure*}[t]
	\centering
	\includegraphics[width=8cm]{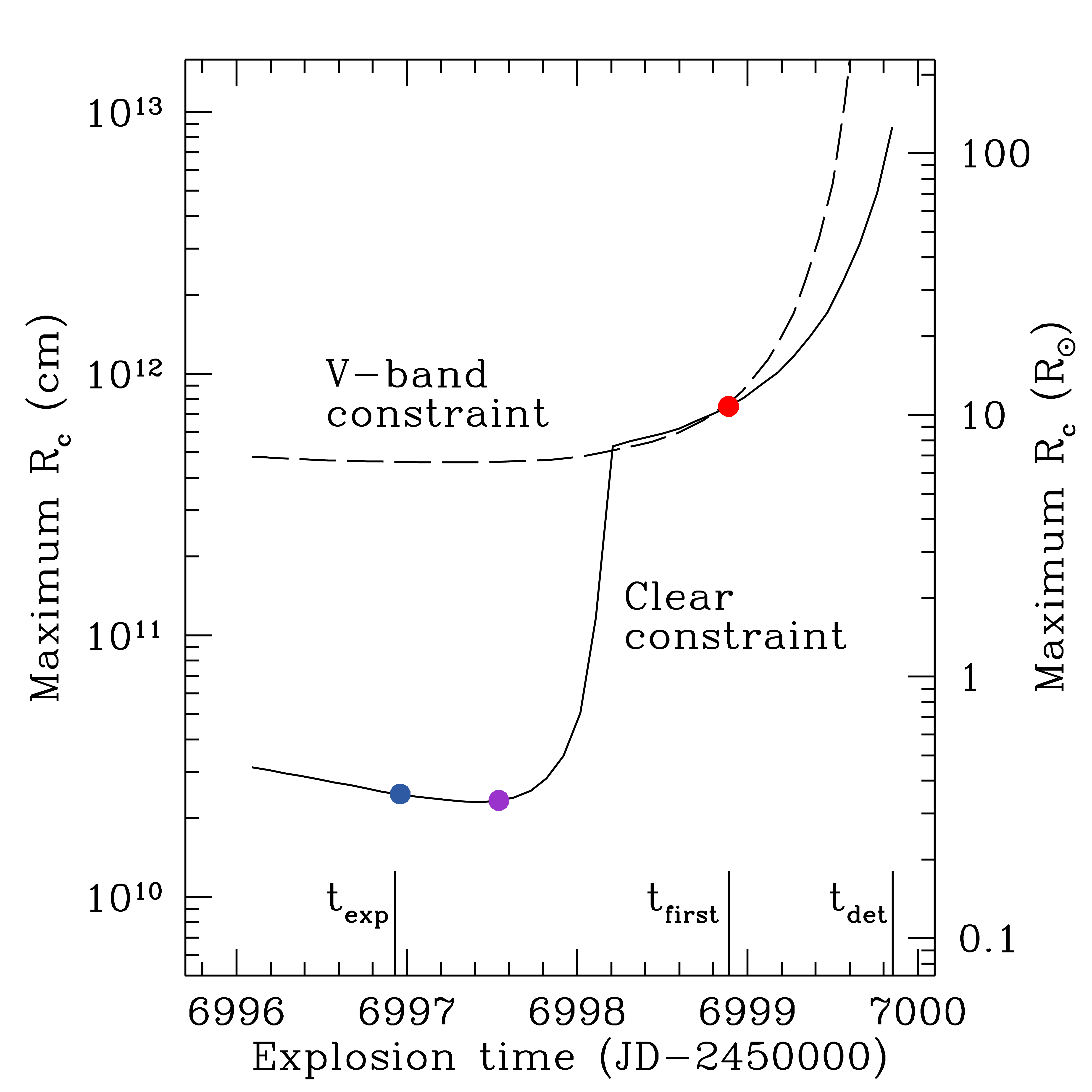}
	\includegraphics[width=8cm]{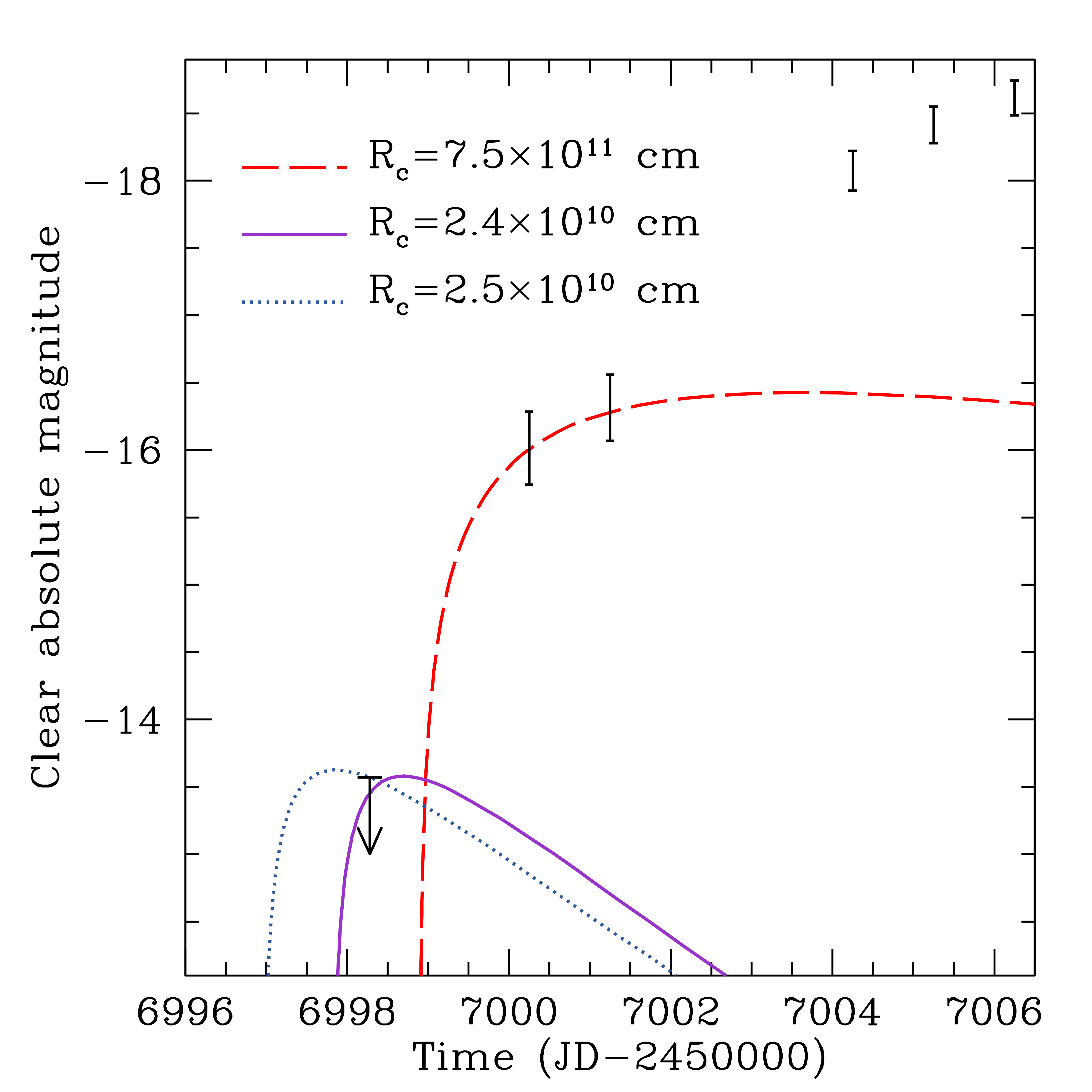}
	
	\caption{Constraints on the companion radius ($R_\textrm{c}$) assuming Roche-lobe overflow and a 1 \msun{} companion.  {\bf Left panel:}  Maximum allowed $R_\textrm{c}$ as a function of explosion time.  The constraints from the ASAS-SN $V$-band and K. Itagaki clear light curves are represented by the dashed and solid lines, respectively.  The first photometric point detecting ASASSN-14lp ($t_{\textrm{det}}$) and the best-fit $t_{\textrm{first}}$ and $t_{\textrm{exp}}$ are indicated.  The purple, blue, and red points indicate the $R_\textrm{c}$ and explosion time of the companion collision light curves shown in the right panel plotted in the same color.  {\bf Right panel:} The early-time, clear light curve of ASASSN-14lp as compared to theoretical light curves for 3 choices of the stellar radius.}
	
	\label{fig:companion}
\end{figure*}

This introduces a weak dependence on the mass ratio of the binary\footnote{The orbital separation is $ \propto (1 + \frac{M_{WD}}{M_S})^{1/3}$, where $M_{WD}$ is the mass of the white dwarf and $M_S$ is the mass of the secondary.}, and we simply assume the primary and companion masses of 1.4 and $1\,\msun$, respectively. The results are summarized in Figure~\ref{fig:companion}, which demonstrates that the constraints on $R_\textrm{c}$ are strongly dependent on the explosion time. If the explosion time is near the $t_{\textrm{first}}$ determined in Section~\ref{sec:lcfit}, a maximum $R_\textrm{c}$ of $\approx31\,\rsun$ is found. On the other hand, if the explosion time is close to $t_{\textrm{exp}}$, then $R_\textrm{c}\lesssim0.35\,\rsun$. These estimates assume a viewing angle of 15 degrees.

Although there are a wide range of potential progenitor scenarios, binary systems with non-degenerate companions can be roughly broken into three main cases: 
1) Systems with a red giant companion, similar to the observed binaries RS Oph and T CrB (e.g., \citealp{hachisu01}), with a large radius greater $> 100 \,\rsun$.  This scenario is difficult to reconcile with ASASSN-14lp for most viewing angles. 
2) Systems with a helium star companion, like the helium-nova system V445 Pup (e.g., \citealp{kato08}), with radii of $\sim 0.3-6\,\rsun$. Even if the explosion time is earlier, this scenario is consistent with our constraints. 
3) Systems with a main-sequence or subgiant companion, similar to U Sco (e.g., \citealp{thoroughgood01}), with radii of $\sim 0.4-4\,\rsun$. A later explosion time does not provide meaningful constraints on this scenario, but a high mass main sequence companion is possible for early explosion times only for unlucky viewing angles. 
Overall, we rule out a red giant companion for all but the most unfavorable viewing angles, but a helium star, a main sequence, or a subgiant companion are all consistent with our observations.

\subsection{Constraints on the Primary's Radius}
\label{sec:primary}

\begin{figure}[t]
	\centering
	\includegraphics[width=8cm]{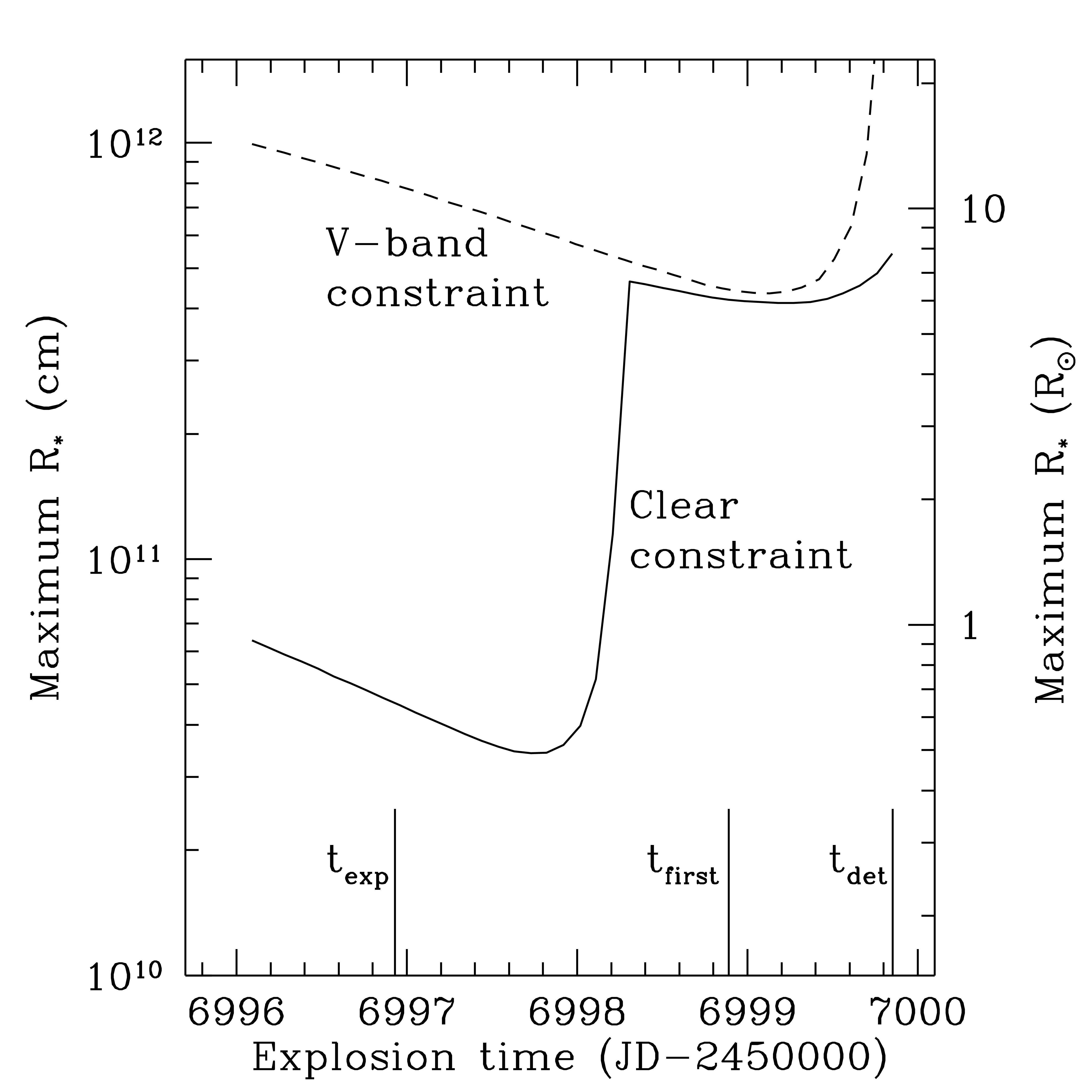}

	\caption{Constraints on the primary star's radius ($R_*$) from a shock-heated envelope as a function of explosion time.  The constraints from the ASAS-SN $V$-band and K. Itagaki clear light curves are represented by the dashed and solid lines, respectively.  The first photometric point detecting ASASSN-14lp ($t_{\textrm{det}}$) and the best-fit $t_{\textrm{first}}$ and $t_{\textrm{exp}}$ are indicated.}
	
	\label{fig:primary}
\end{figure}

Assuming that there is no appreciable emission from the interaction of the SN ejecta and a companion, the first light expected from a \snia{} is the cooling of the expanding, shocked-heated primary star \citep{piro13}.  This emission will, however, be overtaken by radioactive nickel heating after the first few days.  The emission from the cooling of the shock-heated primary is roughly proportional to $R_*$ (\citealp{piro10}).  This fact was utilized by \citet{bloom11} to put tight constraints on the radius of the star that exploded in SN~2011fe. We repeat this exercise for ASASSN-14lp and again consider a variety of explosion times, which we summarize in Figure~\ref{fig:primary}. If the explosion time is near the $t_{\textrm{first}}$ determined in Section~\ref{sec:lcfit}, then a maximum $R_*$ of $\approx6\,\rsun$ is found. On the other hand, if the explosion time is close to $t_{\textrm{exp}}$, then $R_*\lesssim0.6\,\rsun$.  Neither of these constraints is surprising given the general agreement that the progenitor of a \sneia{} is a WD of much smaller size. Nevertheless, since this emission is isotropic, these constraints are firmer than the viewing-angle-dependent constraints from Section~\ref{sec:companion} on interaction with the companion.

\section{Discussion and Summary}
\label{sec:conclusion}

We discovered ASASSN-14lp on 2014 Dec.~9.61, just \discoveryafter{} days after first light.  We announced it to the community less than 5 hours later and we promptly classified it as a \snia{} \citep{thorstensenATEL14}.  Here we added prediscovery photometry, with our first detection less than a day after $t_{\textrm{first}}$.  We also presented ultraviolet through NIR photometric and spectroscopic follow-up data covering the rise and fall of ASASSN-14lp for more than 100 days.  ASASSN-14lp had a broad light curve ($\Delta m_{15}(B) = 0.80  \pm  0.05$), a $B$-band maximum at JD $2457015.82 \pm 0.03$, a rise time of $\powerfittrise$ days, and moderate host--galaxy extinction ($E(B-V)_{\textrm{host}} = 0.33  \pm  0.06$).  Using the light curve of the SN, we determine the distance modulus to the host galaxy, NGC 4666, to be $\mu = 30.8  \pm  0.2$ or a distance of  $14.7 \pm 1.5$ Mpc.  

In a future work, the R$_{V}$ of the host galaxy reddening will be measured using the high quality optical-NIR CSPII data once ASASSN-14lp has faded sufficiently to allow template images to be obtained. Such measurements are of interest because one of the largest systematic uncertainties limiting the precision of \sneia{} as cosmological distance indicator is the uncertainty in the host-galaxy reddening law \citep{mandel11}.  

While there is a Tully-Fisher estimate of the distance (e.g., \citealp{sorce14, springbob07}), a better measurement of the distance to NGC~4666 with HST is needed to add ASASSN-14lp to the surprisingly small calibrating sample (currently 9; \citealp{riess11} and SN~2011fe) of \sneia.  However, NGC~4666 is highly inclined, so crowding and extinction will make a Cepheid distance determination difficult.  Fortunately, the inclination of NGC~4666 is actually an advantage for the tip of the red giant branch method of determining distance because it is easier to probe the halo population with reduced contamination from disk stars \citep{madore98}.  ASASSN-14lp should be added to the ongoing ``Carnegie Hubble Program to Measure $H_0$ to 3\% using Population II'' project (CHP II; \citealp{freedman14hst}) or a similar study.

We used our early-time light curve to constrain JD $t_{\textrm{first}} = \powerfittoall$, implying that $t_{\textrm{rise}} = \powerfittrise$ days and that we discovered ASASSN-14lp at day \discoveryafter.  We then used the early-time spectroscopic data to determine JD $t_{\textrm{exp}} = \velbesterror$ based on the methods of \citet{piro14}.  This is \veldiffbesterror{} days before the $t_{\textrm{first}}$ determined from the light-curve fit, which is similar to the behavior of SN~2009ig \citep{piro14}.  This difference might indicate that there was a significant dark time between the explosion and when ASASSN-14lp began optically brightening or that the explosion of ASASSN-14lp was asymmetrical, but a more detailed study is needed.  

Finally, we used our early-time photometric and spectroscopic data along with our derived light curve properties to place constraints on the progenitor system.  First we constrained $R_* < $ \primaryradius{} depending on the explosion time (independent of viewing angle).  Second, for a viewing angle of 15 degrees, we rule out the presence of a non-degenerate companion with a radius of \radiusconstraint{} or larger.  This range originates from the choice of $t_{\textrm{exp}}$ or $t_{\textrm{first}}$ for the explosion time.  While this is a favorable viewing angle, the strong constraints make it difficult to reconcile ASASSN-14lp with the presence of a red giant companion.

Though ASAS-SN is not designed to find bright SNe Ia as early as possible, the discovery of ASASSN-14lp within $\sim 2$ days after first light is quite encouraging.  Furthermore, during the previous ASAS-SN epoch, within 1 day after first light, ASASSN-14lp was 1.5 mag above the limiting magnitude of ASAS-SN.  Even though we failed to trigger on the transient because of poor weather conditions, this observation demonstrates that sub-day discoveries of bright SNe Ia are possible.  

It is the nearby, bright \sneia{} that have recently allowed us to further our understanding of their progenitor systems.  This is because many of the most constraining measurements can only be performed for these objects: HST pre-explosion imaging (e.g., \citealp{li11, graur12ATEL, graur14, kelly14}), constraints on stripped material from a companion (e.g., \citealp{mattila05, leonard07, shappee13b}), $X$-ray and radio constraints on the presence of circumstellar material (e.g., \citealp{chomiuk12, chomiuk12ATEL, horesh12, pereztorres14}), and constraints on circumstellar material from variable narrow absorption lines (e.g., \citealp{patat07, simon09}).  Moreover, the SNe where multiple constraints have been possible (e.g., SN~2011fe, SN~2012cg, and SN~2014J) have proven to be the most enlightening.

This leads us to question, ``What is the earliest that bright SNe Ia can be found with ASAS-SN or other deeper surveys?''  To address this question, we use the early-time power-law fits of ASASSN-14lp from Table 4 to determine that surveys with limiting magnitudes of $V \sim$ 17 mag and $r \sim$ 20.5 mag would have been capable of detecting ASASSN-14lp 9.5 hours and 1.9 hours after first light, respectively.  In this case, cadence becomes the limiting factor; unless it is more rapid than $\sim$10 hours, deeper surveys do not have an advantage over ASAS-SN. In addition, ASAS-SN has greater sky coverage.

The primary science aim of the ASAS-SN survey is to systematically find nearby supernovae in the entire extragalactic sky and, despite its small (14 cm) lenses, ASAS-SN has become the dominant discover of bright ($V>17$ mag) SNe thus far in 2015 (Shappee et al. 2016 in prep).  Furthermore, the Cassius station in Chile is currently (July 2015) being upgraded from two to four telescopes, to match the Brutus station in Hawaii.  Long term, the goal is to expand ASAS-SN to at least two more stations, which will minimize temporal gaps due to detrimental weather or (rare) technical problems at any given ASAS-SN station.

\acknowledgments

The authors thank Jennifer van Saders, Curtis McCully, Peter Brown, Saurabh Jha, and Louisa Diodato for discussions and encouragement, Toru Yusa for aid in translation, Joe Antognini for reading, and Ryan Foley for providing us with the spectra of SN~2009ig.   We thank Neil Gehrels for approving out ToO requests and the Swift science operation team for performing the observations.  We thank LCOGT and its staff for their continued support of ASAS-SN. We also thank the anonymous referee for his/her helpful comments and suggestions improving this manuscript.

Development of ASAS-SN has been supported by NSF grant AST-0908816 and CCAPP at the Ohio State University. CSK and KZS are supported by NSF grants AST-1515876 and AST-1515927. ASAS-SN is supported in part by Mt.~Cuba Astronomical Foundation.  Operations of the Cassius ASAS-SN station are partially funded from project IC120009 Millennium Institute of Astrophysics (MAS) of the Millennium Science Initiative, Chilean Ministry of Economy.

B.S. is supported by NASA through Hubble Fellowship grant HF-51348.001 awarded by the Space Telescope Science Institute, which is operated by the Association of Universities for Research in Astronomy, Inc., for NASA, under contract NAS 5-26555. TW-SH is supported by the DOE Computational Science Graduate Fellowship, grant number DE-FG02-97ER25308. Support for J.L.P. is in part provided by FONDECYT through the grant 1151445 and by the Ministry of Economy, Development, and Tourism's Millennium Science Initiative through grant IC120009, awarded to The Millennium Institute of Astrophysics, MAS. J.F.B. is supported by NSF grant PHY-1404311.  S.D. is supported by ``the Strategic Priority Research Program- The Emergence of Cosmological Structures'' of the Chinese Academy of Sciences (grant No.~XDB09000000).  E.H. and M.S. are supported by the Danish Agency for Science and Technology and Innovation realized through a Sapere Aude Level 2 grant.  P.R.W. is supported by the Laboratory Directed Research and Development program at LANL.

CSPII is supported by the NSF under grants AST-0306969, AST-0607438, and AST-1008343. 

This paper uses data products produced by the OIR Telescope Data Center, supported by the Smithsonian Astrophysical Observatory.

The Liverpool Telescope is operated on the island of La Palma by Liverpool John Moores University in the Spanish Observatorio del Roque de los Muchachos of the Instituto de Astrofisica de Canarias with financial support from the UK Science and Technology Facilities Council.  

The Nordic Optical Telescope, operated by the Nordic Optical Telescope Scientific Association at the Observatorio del Roque de los Muchachos, La Palma, Spain, of the Instituto de Astrofisica de Canaries.

This paper is based in part on observations obtained through the CNTAC proposal CN2015A-119.

This paper used data obtained with the MODS spectrographs built with funding from NSF grant AST-9987045 and the NSF Telescope System Instrumentation Program (TSIP), with additional funds from the Ohio Board of Regents and the Ohio State University Office of Research. The LBT is an international collaboration among institutions in the United States, Italy and Germany. 

This paper includes data gathered with the 6.5 meter Magellan Telescopes located at Las Campanas Observatory, Chile.

IRAF is distributed by the National Optical Astronomy Observatory, which is operated by the Association of Universities for Research in Astronomy (AURA) under a cooperative agreement with the National Science Foundation.

Funding for the SDSS and SDSS-II has been provided by the Alfred P. Sloan Foundation, the Participating Institutions, the National Science Foundation, the U.S. Department of Energy, the National Aeronautics and Space Administration, the Japanese Monbukagakusho, the Max Planck Society, and the Higher Education Funding Council for England. The SDSS Web Site is http://www.sdss.org/.  

This research has made use of data from the AAVSO Photometric All Sky Survey, whose funding has been provided by the Robert Martin Ayers Sciences Fund.

This research has made use of the NASA/IPAC Extragalactic Database (NED) which is operated by the Jet Propulsion Laboratory, California Institute of Technology, under contract with the National Aeronautics and Space Administration.  

This research has also made use of the XRT Data Analysis Software (XRTDAS) developed under the responsibility of the ASI Science center (ASDC), Italy.


\bibliographystyle{apj}

\end{document}